# Topological beaming of light: Proof-of-concept experiment

Yu Sung Choi[1,*], Ki Young Lee[1,*], Soo-Chan An[1], Minchul Jang[2], Youngjae Kim[2], Seungjin Yoon[3], Seung Han Shin[1], and Jae Woong Yoon[1,†]

[1]Department of Physics, Hanyang University, Seoul, 133-791, Korea
[2]Convergence Technology Division, Korea Advanced Nano Fab Center, Suwon 16229, Korea
[3]Joint Quantum Institute, University of Maryland, College Park, MD 20742, USA
*These authors contributed equally to this work

†Correspondence should be addressed to yoonjw@hanyang.ac.kr

**Beam shaping in nanophotonic systems remains a challenge due to the reliance on complex heuristic optimization procedures. In this work, we experimentally demonstrate a novel approach to topological beam shaping using Jackiw-Rebbi states in metasurfaces. By fabricating thin-film dielectric structures with engineered Dirac-mass distributions, we create domain walls that allow precise control over beam profiles. We observe the emergence of Jackiw-Rebbi states and confirm their localized characteristics. Notably, we achieve a flat-top beam profile by carefully tailoring the Dirac-mass distribution, highlighting the potential of this method for customized beam shaping. This experimental realization establishes our approach as a new mechanism for beam control, rooted in topological physics, and offers an efficient strategy for nanophotonic design.**

## Introduction

Precise control and shaping of light beams are crucial for a wide range of applications, including laser machining, laser therapy, optical communications, and emerging quantum technologies[1-9]. Traditionally, beam shaping has relied on optical elements such as refractive and diffractive optical elements (DOEs)[10-14] and spatial light modulators (SLMs)[15-17]. Although these methods are effective, they generally involve heuristic optimization algorithms, substantially limiting their adaptability particularly for intricate nanophotonic structures.

Recent advances in topological photonics have opened new avenues for manipulating light in unprecedented ways[18-21]. Among various intriguing effects, the Jackiw-Rebbi (JR) soliton, a zero-energy solution for a domain-wall Dirac equation, attracts considerable interest because of its significance in fundamental physics of solid-state systems and topological photonic device applications[22-33]. Toward this end, topological beam-shaping method was theoretically proposed as a promising design principle[34]. In this method, a photonic analogy of the Jackiw-Rebbi (JR) state and its new degree of control freedom enables a remarkably efficient beam shaping in a systematic manner that does not require tedious optimization steps.

Here, we present the first experimental demonstration of the topological beam-shaping principle. Utilizing lossless dielectric media in the optical telecommunications band, we fabricate a metasurface structure that its internal resonance states emit a beam of light with arbitrarily designed shape in principle. The fabricated devices reveal key spectral and spatial properties of the prescribed topological state and indeed produce the desired flat-top beams in experiments. We discuss the inherent limitations of our experimental results and propose a potential solution based on monolithic geometry control of the unit-cell design.

## Results

### Theory

The leakage-radiation beam shaping here is schematically illustrated in Fig. 1a. A waveguide grating with carefully designed unit-cell structure supports a leaky guided modes with a prescribed standing-wave envelope profile. A desired beam of leakage radiation is emitted by the first-order diffraction from the guided-mode fields as it transfers the standing-wave envelop profile in the guided mode to the emitted beam.

Obtaining an appropriate waveguide-grating design for a certain desired standing-wave envelope in the guided-mode field is nontrivial and generally involves trial-solution-based numerical optimization algorithms in the similar manner as conventional beam-shaping diffractive optical elements. Intriguingly, our previous work[34] theoretically suggests a remarkably simple rule which can be expressed in a closed-form equation as

$$\kappa = -c\frac{1}{f}\frac{df}{dx}, \quad (1)$$

where $\kappa$ is the second-order Bragg-reflection rate of the guided mode, $f$ is a desired beam profile, $c$ is speed of light in vacuum, and $x$ is position along the grating periodicity. The rule by Eq. 1 applies to a specific resonance state occurring in the middle of the second stop band for the guided mode.

For a waveguide grating structure with a single ridge per unit cell, the second-order Bragg-reflection rate takes an expression

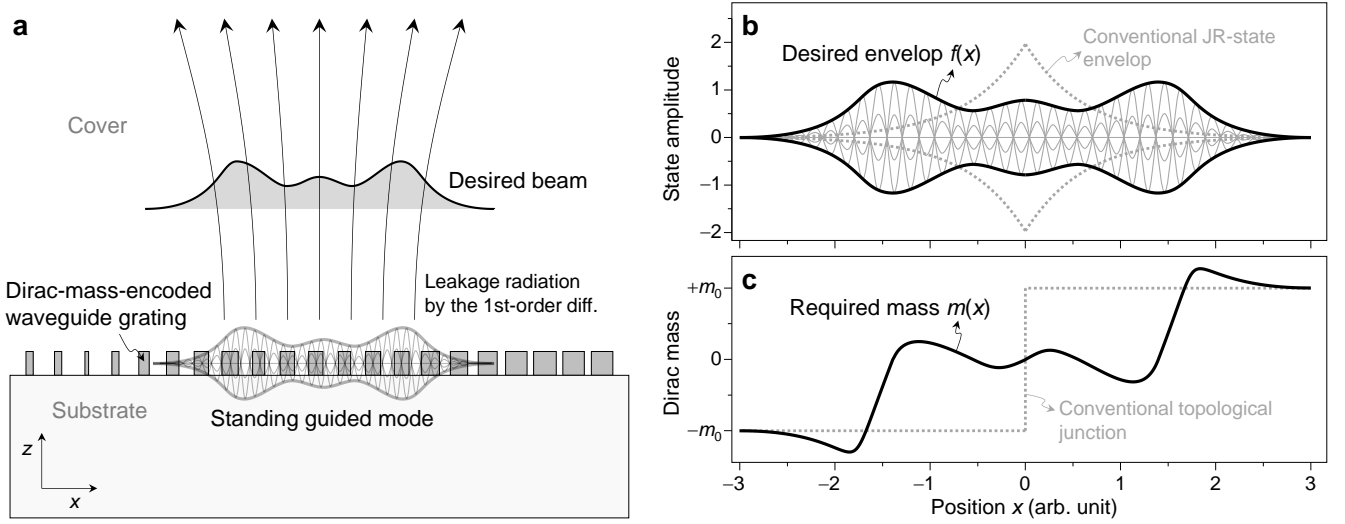

**Fig. 1 Principles of topological beam shaping and Jackiw-Rebbi state engineering in metasurfaces. a** Schematic representation of leakage-radiation beam shaping with engineered Dirac mass distribution. The guided-mode resonance (GMR) with a tailored standing-wave envelope emits a desired beam profile through first-order diffraction. **b** Characteristic bi-exponential envelope profile $f(x)$ of the Jackiw-Rebbi (JR) state at a simple topological junction (gray dotted curve). **c** Corresponding spatially-varying Dirac mass distribution $m(x)$ that generates the JR state profile in **b**.

$$\kappa = \left[ D_1 F^2 \Delta\varepsilon^2 \operatorname{sinc}^2(F) - D_2 F \Delta\varepsilon \operatorname{sinc}(2F) \right] \omega_0, \quad (2)$$

where $\Delta\varepsilon$ is dielectric-constant difference between the high-index and low-index parts of the grating, fill factor $F$ is relative width of the grating ridge to the period, and $\omega_0$ is frequency of the second-order Bragg reflection for the guided mode. $D_m$ is diffraction-strength constant which is determined by the following relations.

$$D_1 = \frac{n_p}{2n_g} \iint_{\text{Grating layer}} dz\, dz'\, k_0^2\, u^*(z) G(z,z') u(z'), \quad (3)$$

$$D_2 = \frac{n_p}{2n_g} \int_{\text{Grating layer}} dz\, |u(z)|^2, \quad (4)$$

where $n_p$ and $n_g$ denote phase and group speed indices, respectively, $k_0$ is vacuum wave number, $u(z)$ is normalized wave function of the guided mode, and $G(z,z')$ is 1D Green's function for an effective 1D structure that the grating layer is replaced by a homogeneous effective medium. We note that $D_1$ denotes strength of two consecutive first-order diffractions of the guided mode in the grating layer while $D_2$ indicate strength of the single second-order diffraction. See[25,34] for details of the theory leading to these relations. Combining Eqs. 1 and 2, one can readily obtain beam-shaping GMR structure with a simple mapping between desired envelope profile $f(x)$ and fill-factor distribution $F(x)$ such that

$$\left[ D_1 F^2 \Delta\varepsilon \operatorname{sinc}^2(F) - D_2 F \operatorname{sinc}(2F) \right] = -\Delta\varepsilon^{-1} \frac{1}{f} \frac{df}{k_0 dx}. \quad (5)$$

Let us briefly explain how this remarkably simple method of leakage-radiation beam shaping works. It is based on wave-kinematic analogy of GMRs to 1D Dirac fermions governed by the well-known Hamiltonian

$$\mathbf{H}_D = mc^2 \boldsymbol{\sigma}_3 - pc\, \boldsymbol{\sigma}_1 = \begin{bmatrix} mc^2 & -pc \\ -pc & -mc^2 \end{bmatrix}. \quad (6)$$

Here, $m$ is Dirac mass, $p$ is linear momentum, and $\boldsymbol{\sigma}_j$ is the Pauli matrix. For a 1D topological junction at $x = x_0$ where the space is distinguished such that $m(x < x_0) = -m_0$ and $m(x > x_0) = +m_0$ for $m_0 > 0$, a zero-energy eigenstate taking expressions

$$|0\rangle_D = \frac{1}{\sqrt{2}} \begin{bmatrix} 1 \\ i \end{bmatrix} f(x), \quad (7)$$

$$f(x) = \exp\left[ -\frac{c}{\hbar} \int_0^x m(x') dx' \right] = \exp\left( -\frac{c}{\hbar} m_0 |x| \right) \quad (8)$$

exists. This eigenstate at the center of the bandgap is referred to as Jackiw-Rebbi (JR) soliton[22]. In Figs. 1b and 1c, the gray-dotted curves schematically indicate typical bi-exponential envelope profile $f(x)$ and corresponding Dirac mass distribution $m(x)$ for this simple topological-junction system, respectively.

Equation 8 for the JR-state solution immediately imply that we can obtain arbitrarily shaped $f(x)$ beyond the simple bi-exponential profile, as schematically illustrated in Figs. 1b and 1c. If we extend the piece-wise constant $m(x)$ distribution to an arbitrary continuous function containing at least one or more junction points at which sign of $m$ changes, $f(x)$ can take

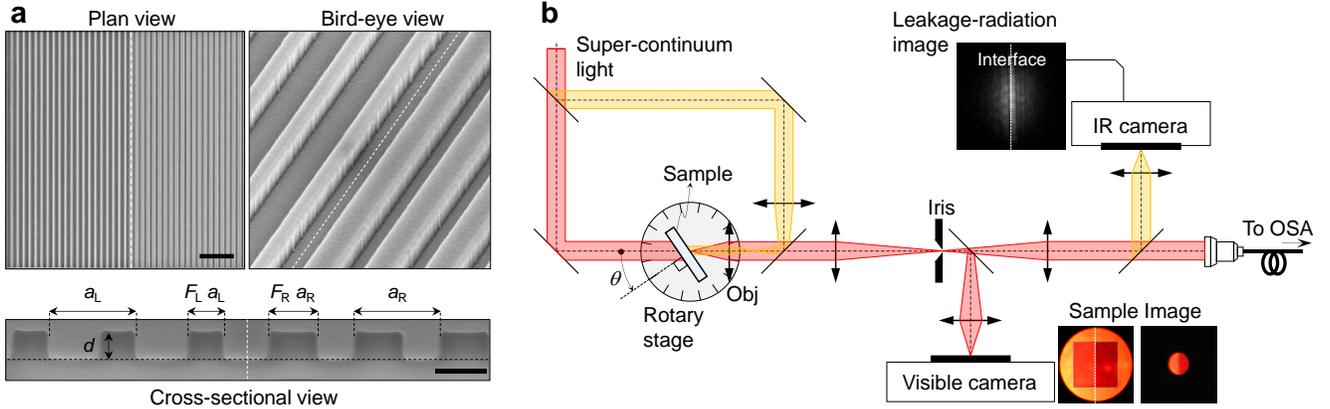

**Fig. 2 Experimental realization of topological Jackiw-Rebbi state. a** Focused ion beam image of the fabricated topological junction device from plan (scale bar, 5 μm), bird-eye and cross-sectional (scale bar, 500 nm) views. **b** Schematic of the angle-resolved confocal microscopy setup. Obj, objective lens; TL, tube lens; OSA, optical spectrum analyzer. Inset images display the sample (obtained from visible camera) and radiation pattern (from IR camera), with white dashed lines marking the topological interface.

any arbitrary shape in principle. In this respect, Eq. 8 can be alternatively expressed as

$$\frac{c}{\hbar}m(x) = -\frac{1}{f}\frac{df}{dx}, \qquad (9)$$

which inversely determines $m(x)$ distribution for certain desired envelope profile $f(x)$.

Equation 9 for 1D Dirac fermions is directly applicable to GMR states because they are connected through a unitary transformation that preserves the eigen-system structures and pertaining wave-kinematic properties. Following the coupled-mode theory by Kazarinov and Henry[35-38], a GMR state is described by an eigenvalue problem $\mathbf{H}_G|\psi\rangle_G = \omega_G|\psi\rangle_G$ with the Hamiltonian taking an expression

$$\mathbf{H}_G = kv_g\boldsymbol{\sigma}_3 + \kappa\boldsymbol{\sigma}_1 = \begin{bmatrix} kv_g & \kappa \\ \kappa & -kv_g \end{bmatrix}, \qquad (10)$$

where $|\psi\rangle_G = [\psi_R \; \psi_L]^T$ is eigenvector with $\psi_R$ and $\psi_L$ denoting amplitudes of guided modes propagating towards right (R) and left (L) in $x$-axis, respectively, $k$ is Bloch wavevector, $\omega_G$ is frequency eigenvalue indicating relative frequency from second-order Bragg reflection frequency $\omega_0$, $\kappa$ is second-order Bragg-reflection rate for the guided mode as already defined in Eqs. 1 and 2, $v_g$ is group speed of the guided mode, and $\boldsymbol{\sigma}_n$ is Pauli matrix. Guided-mode Hamiltonian $\mathbf{H}_G$ in Eq. 10 and 1D Dirac Hamiltonian $\mathbf{H}_D$ in Eq. 6 are related to each other through a unitary transformation such that

$$\mathbf{U}\mathbf{H}_D\mathbf{U}^\dagger = \hbar\mathbf{H}_G, \qquad (11)$$

$$\mathbf{U} = \frac{1}{\sqrt{2}}(\boldsymbol{\sigma}_0 - i\boldsymbol{\sigma}_2) = \frac{1}{\sqrt{2}}\begin{bmatrix} 1 & -1 \\ 1 & 1 \end{bmatrix} \qquad (12)$$

with parametric correspondences

$$p = \hbar k, \; c = v_g, \text{ and } mc^2 = \hbar\kappa. \qquad (13)$$

Therefore, the JR state $|JR\rangle_G$ for guided modes is derived from the JR state $|0\rangle_D$ in Eq. 7 for 1D Dirac fermions as

$$|JR\rangle_G = \mathbf{U}|0\rangle_D = \frac{1}{\sqrt{2}}\begin{bmatrix} e^{-i\pi/4} \\ e^{+i\pi/4} \end{bmatrix}f(x), \qquad (14)$$

$$f(x) = \exp\left[-\frac{1}{c}\int_0^x \kappa(x')dx'\right]. \qquad (15)$$

Finally, Eq. 1 for the leakage-radiation beam shaping is followed by taking the first-order derivative of Eq. 15 with respect to $x$.

**Experiment: Basic spectral properties**
To verify our theoretical proposal in experiments, we fabricate GMR-structure samples using standard e-beam lithography. Briefly, a silicon-nitride (SiN) layer with thickness of 300 nm is first deposited on the quartz wafer as the slab waveguide layer. Then, the SiN grating patterns are produced using e-beam lithography and etching process. A 2-μm-thick $SiO_2$ cover layer is deposited. The focused ion beam (FIB) images for a topological junction sample are shown in Fig. 2a. The left and right sides of the junction have different fill-factors $F$ and periods $a$. In this case, we chose $F_L = 0.3$, $a_L = 880$ nm for the left side and $F_R = 0.6$, $a_R = 840$ nm for the right side. See Materials and Methods for details of the fabrication steps and conditions.

In spectral measurement of the sample for the desired topological properties, we use a microscopic angle-resolved spectrum analysis set up, as schematically illustrated in Fig. 2b. The set up acquires angle-resolved transmittance spectrum at a designated microscopic spot on the sample, enabling position-dependent spectrum analysis. A collimated supercontinuum light beam (highlighted in red) is incident on the sample at angle $\theta$. The transmitted beam is collected by the objective lens. The tube lens forms a magnified image of the sample. An iris-diaphragm at the magnified-image plane designates the measurement spot on the sample. The image of the selected spot is acquired with the visible-image sensor on the bottom in order to conveniently adjust size and location of the measurement spot. Meanwhile, part of the transmitted light is coupled to a fiber connected to an optical spectrum analyzer. The parts associated with the beams highlighted in yellow are for leakage-radiation beam analysis that will be explained in the next section.

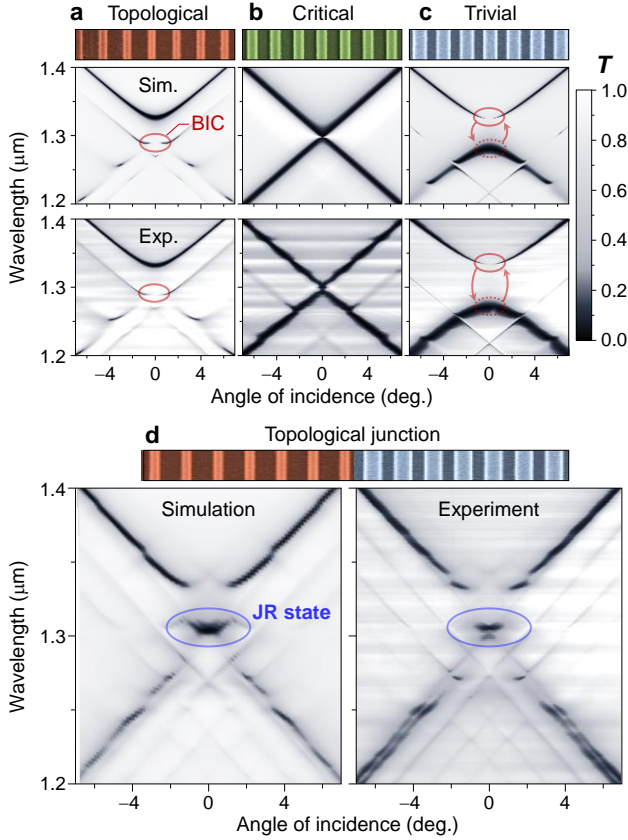

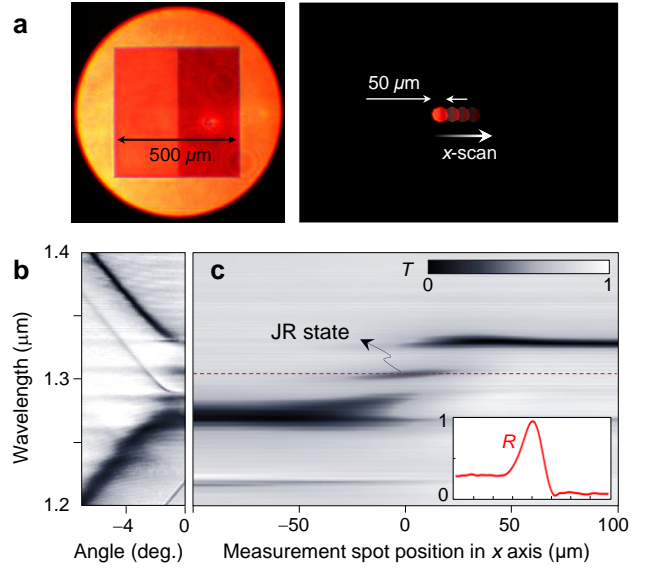

**Fig. 3 Angle-resolved transmission spectra demonstrating topological phase transition and emergence of the Jackiw-Rebbi state.** **a** Topological phase, $F = 0.3$, $a = 880$ nm. **b** Critical phase, $F = 0.44$, $a = 860$ nm. **c** Trivial phase, $F = 0.6$, $a = 840$ nm. **d** Topological junction sample, revealing a localized JR state within the bandgap.

**Fig. 4 Spatial characteristics of the Jackiw-Rebbi state.** **a** Visible camera images of the sample and measurement area. **b** Angle-resolved transmission spectrum of the junction structure. **c** Local resonance-spectrum under normal incidence with inset showing reflectance profile at the JR resonance wavelength (red dashed).

Angle-resolved transmittance spectra of our fabricated samples reveal definite evidence for the topological phase transition and emergence of the photonic JR state. In Fig. 3, we show experimental spectra in excellent agreement with the rigorous numerical simulation by the finite-element method (FEM). Figures 3a ~ 3c experimentally confirm the topological phase transition with the fill-factor change. A characteristic feature of the topological phase transition is a spectral flip of dark and bright resonance band edges. We indicate the dark resonance, often referred to as bound state in the continuum (BIC), with red ellipse in Fig 3a for the topological phase for $F = 0.3$. The BIC is at the lower-wavelength band edge therein. As $F$ increases to 0.44, the GMR is at the topologically critical phase and Dirac-point-like resonance-band crossing appears, as shown in Fig. 3b. As $F$ increases further, the band gap opens again and the BIC is now at the upper-wavelength band edge, as shown in Fig. 3c for $F = 0.6$.

Of particular interest in our study here is the JR-state resonance for topological junction structures. In Fig. 3d, we show the spectrum for the junction sample consisting of two topologically distinguished structures in Figs. 3a and 3c. The experimental spectrum contains a clear signature of the JR-state resonance at the center (~ 1.3 μm) of the band gap again in excellent agreement with the numerical simulation. In further detail, this JR-state resonance corresponds to the case illustrated in Figs. 1b and 1c for the conventional JR-state envelope. Thereby, it takes a bi-exponential decaying envelope profile localized at the junction interface. Although the observed JR-state feature in the spectral domain is a definite evidence, more direct observation of its localization property should be further investigated to completely reveal its unique characteristics in experiment.

Consequently, we perform local spectral analysis by taking measurement aperture size at its our technical minimum ~ 50 μm, as shown in Fig. 4a. We acquire normal-incidence transmittance spectrum while the 50-μm-wide spot scans its x-position across the topological-junction interface. The measured spectra are summarized in Figs. 4b and 4c. In Fig. 4b, we show the angle-resolved spectrum when the spot is exactly at the junction interface for reference. The JR-state resonance feature is at wavelength 1.3 μm and incident angle 0. The $x$-scan normal-incidence spectrum in Fig. 4c confirms that the JR-state resonance is localized to the junction interface as the resonance feature is excited only when the spot is in close vicinity of the junction interface. This localization property is in stark contrast to the spatially uniform band-edge resonance features at 1.27 and 1.33 μm.

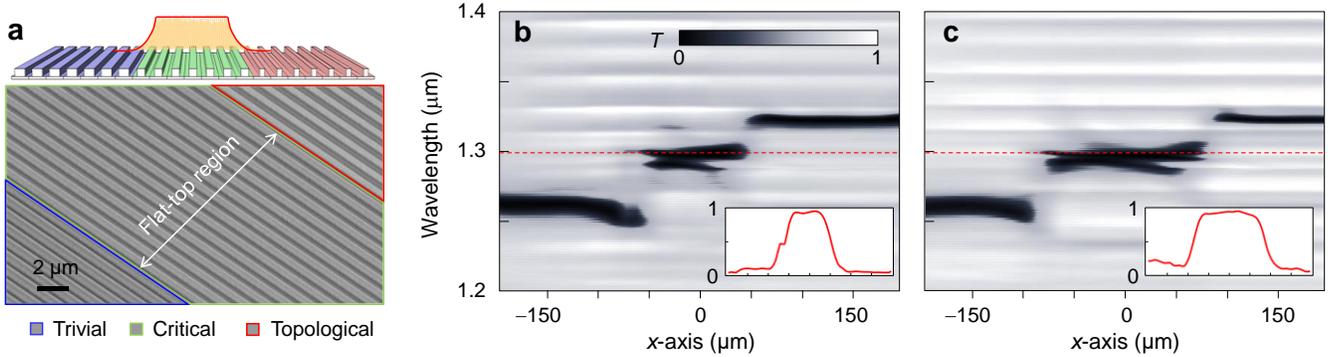

**Fig. 5 Experimental realization of flat-top beam shaping. a** Schematic overview of the structure designed for flat-top beam profile generation, illustrating trivial (blue), critical (green), and topological (red) phases. Scale bar, 2 μm. (**b** and **c**) Local transmission spectra for the junction structure with flat-top regions of widths 85 μm and 127.5 μm, respectively. Insets display the reflectance spectra at the JR state resonance wavelength (red dashed line, 1.3 μm), demonstrating the achieved flat-top beam profiles.

Full-width at half maximum (FWHM) of the JR-state feature in the $x$-scan data is 29 μm, which is substantially larger than the theoretical JR-state envelope width ∼ 12 μm. This width broadening is attributed to boxcar-average blurring due to the 50-μm-wide aperture. Although significant blurring effect exists in the $x$-scan spectral data, it approximates the JR-state envelope profile as indicated in the inset of Fig. 4c. Therein, we indicate the $x$-scan reflectance spectrum at the JR-state resonance-center wavelength (red-dashed line).

**Experiment: Leakage-radiation beam shaping**

As discussed in the theory section, the leakage-radiation distribution from the guided-mode JR state can be adaptably shaped into any desired arbitrary form by the Bragg-reflection-rate mapping. In this section, we investigate this intriguing possibility in experiment.

We fabricate a specific junction structure that creates a flat-top beam, as shown in Fig. 5a. Such junction consists of three piece-wise-constant regions of $m(x < -w/2) = +m_0$ (trivial phase), $m(-w/2 \leq x \leq +w/2) = 0$ (critical phase), and $m(x > +w/2) = -m_0$ (topological phase), respectively. These regions are indicated by blue, green, and red highlights in Fig. 5a for the design schematic on top and actual fabricated sample on the bottom.

For fabricated samples with $w = 85$ and 127.5 μm, we perform the local resonance-spectrum analysis identical to Fig. 4b, as shown in Figs. 5b and 5c, respectively. We see clear JR-state resonance features at 1.3 μm wavelength, which are elongated in $x$-axis over domains consistent with applied $w$ values. The reflectance spectra at 1.3 μm wavelength are provided in the insets and they show favorably flat-top profiles as expected. Therefore, we presume that the JR-states in these samples have flat-top envelope profiles at the desired width values.

Now, we observe leakage-radiation distribution from the JR-state resonance. In our passive GMR structure case, directly observing pure leakage radiation is quite challenging because there are no emissive or florescent elements that excite the resonance state in the absence of light incidence at the resonance wavelength. Consequently, we take a passive mode-filtering technique. A sufficiently wide and collimated light beam is incident on the sample at $\theta = 0$ and reflected beam distribution on $x$-$z$ plane at the resonance wavelength is acquired by an IR camera scanning its object plane from the sample surface ($z = 0$) to a certain prescribed distance ($z = L$). This measurement is done with the set-up in Fig. 2b by incorporating the additional beam paths highlighted in yellow and their pertaining components. We take two reflected beam distributions $U_s(x, z)$ and $U_{\text{ref}}(x, z)$ separately from the topological-junction sample and unpatterned reference area, respectively. Finally, we take $U_{\text{LR}}(x, z) = U_s(x, z) - U_{\text{ref}}(x, z)$ as an approximate leakage-radiation distribution. This method is based on general property of guided-mode resonances that the reflected light $U_s$ can be decomposed by dominant leakage radiation $U_{\text{LR}}$ part from the resonance state and weak non-resonant zero-order reflection $U_0$ part. Since $U_{\text{ref}}$ from the unpatterned area approximates $U_0$ in our moderate index-contrast structure, we can infer $U_{\text{LR}}$ as $U_{\text{LR}}(x, z) = U_s(x, z) - U_0(x, z) \approx U_s(x, z) - U_{\text{ref}}(x, z)$.

The measurement result is summarized in Fig. 6. We present experimentally inferred $U_{\text{LR}}(x, 0 \leq z \leq 200\ \mu\text{m})$ over a domain $|x| \leq 200$ μm and $0 \leq z \leq 200$ μm for four selected $m(x)$ conditions of $w = 0$, 85 μm, 127.5 μm, and 500 μm under Gaussian beam incidence with diameter 80 μm. For each case, we compare the experimental beam profiles on four selected planes at $z = 5, 50, 100, 150$ μm with rigorous numerical calculation due to the FEM. We confirm good agreement between the experiment and simulation for all four $m(x)$ conditions. In particular, the leakage-radiation beam width increases with increasing junction-domain width $w$ from 0 to 275.5 μm, as seen from Figs. 6a ~ 6c. The case for $w = 275.5$ μm in Fig. 6c exemplifies the strong effect of the JR-state resonance because flat-top-like beam profile is observed even though the junction width is considerably wider than 80 μm for the incident beam width.

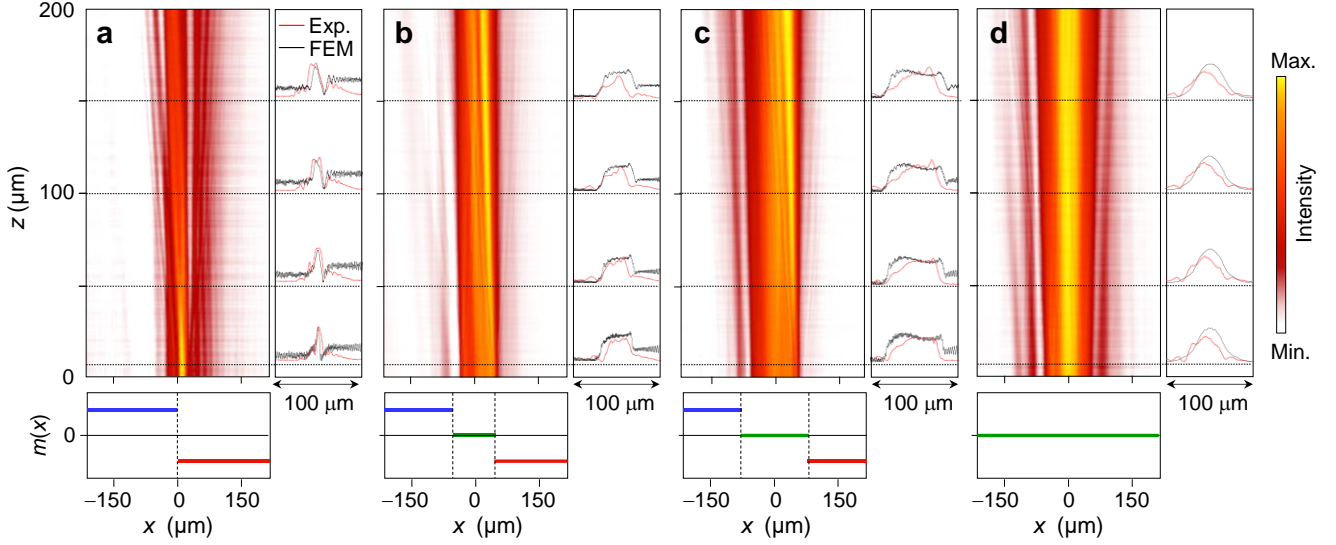

**Fig. 6 Measured leakage-radiation distributions $U_{LR}(x, z)$ for various Dirac mass configurations.** (**a–c**) Topological junction structure with critical phase widths $w$ of 0, 85 μm, 127.5 μm, respectively. **d** Conventional GMR structure with constant Dirac mass ($w = 500$ μm). Left panels show 2D distributions over $|x| \leq 200$ μm and $0 \leq z \leq 200$ μm. Right graphs compare experimental beam profiles (red lines) with FEM simulations (black lines). Bottom schematics illustrate corresponding Dirac mass distributions $m(x)$: blue for positive (trivial phase), green for zero (critical phase), and red for negative (topological phase).

The flat-top-like beam profile is clearly due to the characteristic effect of the topological junction because a conventional GMR structure does not have such property. In Fig. 6d for $w = 500$ μm, the leakage-radiation beam takes a Gaussian profile which is simply identical to the incident beam. In this case, the entire sample domain with footprint width 500 μm has a constant Dirac mass at $m(x) = 0$ with no junction structure.

## Discussion

Although the results in Fig. 6 confirm remarkable effect of the topological junction structure on the leakage-radiation beam properties, neither the experimental nor numerical beam shapes precisely show the desired flat-top profiles. This issue should be properly addressed because accurate generation of a desired beam profile is crucial in consideration of practical applications.

We attribute the observed inaccuracy to inhomogeneity of first-order diffraction amplitude across the junction region. In Fig. 6, an apparent error is the asymmetric beam shape which is certainly not a feature of the JR-state envelope. Assuming that JR-state envelope $f(x)$ is symmetric with respect to the junction center at $x = 0$ as intended from the structure design, the only factor that can possibly produce the strong asymmetry in the emitted beam is uneven distribution of first-order diffraction amplitude. Although it was not pointed out in our original theoretical proposal[34], uniform first-order diffraction amplitude is a necessary requirement for the precise beam shaping.

In further detail, the leakage radiation wavefunction $L_k(z)$ for single in-plane wavevector $(k)$ component is described by the following expression.

$$L_k(z) = \varepsilon_1 (\psi_L + \psi_R) W_k(z), \quad (16)$$

$$W_k(z) = -k_0^2 \int_{\substack{\text{Grating}\\\text{layer}}} dz' G(z, z') u(z'), \quad (17)$$

where $\varepsilon_1$ is first-order harmonic amplitude of the periodic dielectric function in the grating layer and $W_k$ is leakage radiation wave function in $z$ axis. Combining Eq. 16 with the JR-state solution in Eqs. 14 and 15, we obtain an approximate leakage-radiation wave function $L_{JR}$ as

$$L_{JR}(x, z) \approx \varepsilon_1(x) f(x) W_k(z). \quad (18)$$

Although a complete description of $L_{JR}$ with non-uniform excitation of guided modes requires a weighted superposition of $L_k$'s for different $k$ components taking Fourier transform $F(k)$ of $f(x)$ as the weight, Eq. 18 should provide a reasonable approximation for slowly-varying envelope cases. As clearly revealed in Eq. 18, the leakage-radiation profile is $\varepsilon_1(x) f(x)$, not $f(x)$. Consequently, local first-order Fourier coefficient $\varepsilon_1(x)$ has to be constant in $x$ in order to produce an exact replica of $f(x)$ in beam profile $L_{JR}$.

Unfortunately, keeping $\varepsilon_1(x)$ constant for the precise beam shaping seems impossible when we consider that certain non-uniform $\varepsilon_1(x)$ distribution is inevitable for synthesizing required $\kappa(x)$ distribution according to Eqs. 2 and 5. Equation 2 for the second-order Bragg-reflection amplitude is alternatively expressed in terms of Fourier transform $\varepsilon_m$ of the dielectric function in the grating layer as

$$\kappa = \left(D_1 \varepsilon_1^2 - D_2 \varepsilon_2\right)\omega_0. \tag{19}$$

This relation implies that we have to be able to change $\varepsilon_2$ without altering $\varepsilon_1$ for a required $\kappa$ distribution with constant $\varepsilon_1$. For single-ridge unit-cell structures, fill factor $F$ is the only factor that can be used to tune $\kappa$ and it simultaneously changes both $\varepsilon_1$ and $\varepsilon_2$ because $\varepsilon_1 = F\Delta\varepsilon\,\text{sinc}(F)$ and $\varepsilon_2 = F\Delta\varepsilon\,\text{sinc}(2F)$. Therefore, the precise beam shaping with a single-ridge unit-cell structure is in principle impossible, significantly restricting application potential of our Dirac-mass-control approach.

However, changing $\varepsilon_2$ without altering $\varepsilon_1$ may become readily possible if we break the grating ridge into multiple parts within the unit cell. In such compound-grating structures, relative positions of multiple parts can be used as extra degrees of design freedom, which allow various schemes for tuning $\varepsilon_2$ under the constraint of constant $\varepsilon_1$.

In conclusion, we have successfully demonstrated the topological beam-shaping approach through experimental validation. By leveraging the wave-kinematic analogy between guided modes and 1D Dirac fermions, we designed and fabricated a topological-junction structure that supports a guided-mode standing-wave field and its leakage radiation with the desired envelope distribution. Experimental data, obtained through angle-resolved local spectrum analysis and passive beam-profiling method, show excellent agreement with theoretical predictions. However, the results also highlight a limitation in the precision of the generated beam shape. We identify a primary cause of this limitation and propose a potential solution involving compound unit-cell structures, which avoids the need for additional materials or fabrication steps. The compound unit-cell structure is particularly promising for future research, as it not only has the potential to enhance beam shape accuracy but also may facilitate the creation of higher-dimensional topological phases with greater design flexibility. These advancements are crucial for both practical applications and fundamental studies in topological physics.

## Materials and methods
### Numerical simulation details
All numerical simulations here are performed using a commercial FEM solver (COMSOL Multiphysics 5.5). Two-dimensional models (in the x-y plane) are created to simulate one-dimensional photonic lattices with perfectly matched layers along the y-direction. The scattering and periodic (Floquet) boundary conditions are applied along the x-direction and y-direction, respectively. The junction structure was simulated for guided mode resonance lattices with trivial and topological phases, each composed of 20-unit cells in a supercell arrangement. The transmission spectra in Fig. 3 are calculated for TE-polarized plane waves at incident angles ranging from −7° to 7°, with measurements taken from the output port.

### Fabrication
The sample is fabricated on a double-side polished quartz wafer with a silicon nitride layer thickness of 300 nm, and a silica substrate of ~ 500 μm. The step-by-step fabrication process is illustrated in Supplementary Fig. S1. A 300 nm thick SiN film is thermally deposited using plasma-enhanced chemical vapor deposition (PECVD). A layer of ZEP520A e-beam resist (300 nm thick) is spin-coated on the SiN layer. The one-dimensional photonic lattice pattern is then defined using electron-beam lithography (JBX 9300-FS). The photonic patterns are defined in the photo-resist layer, which is then developed in pentyl acetate (ZED-N50) for 90 sec. The SiN layer with the resist pattern is etched in a inductively coupled plasma-reactive ion etcher with $SF_6$ + $O_2$ gas mixture. After residual PR removal, a 2 μm thick $SiO_2$ cover is deposited using PECVD. Finally, the fabricated sample on the 4" quartz wafer is cleaved into 2 cm × 2 cm chips and cleaned. See Supplementary Materials for further details of fabrication steps and conditions.

**Acknowledgements**

This research was supported by the Leader Researcher Program (NRF-2019R1A3B2068083), The National Research Facilities and Equipment Center (NFEC) at the Ministry of Science and ICT (Support from the supporting project for advancement of leading research facilities (PG2023003-03) and the quantum computing technology development program of the Quantum Information Research Support Center, funded through the National research foundation of Korea (2020M3H3A1110365).

**Author contributions**
K.Y.L., Y.S.C., and J.W.Y. conceived the original concept. Y.S.C. and K.Y.L. designed the samples and performed numerical analysis. Y.S.C, M.J., and Y.K. fabricated the samples. Y.S.C., S.C.A., S.Y., and S.H.S performed the measurement and data analysis. J.W.Y. supervised this research. Y.S.C. and J.W.Y. wrote the manuscript. All authors discussed the result.

**Conflict of interest**
The authors declare that they have no conflict of interest.

**Data availability**
All data needed to evaluate the conclusions in the paper are present in the paper and/or the Supplementary Materials.

# Supplementary Information for: Topological beaming of light: Proof-of-concept experiment

Yu Sung Choi[1,*], Ki Young Lee[1,*], Soo-Chan An[1], Minchul Jang[2], Youngjae Kim[2], Seungjin Yoon[3], Seung Han Shin[1], and Jae Woong Yoon[1,†]

[1]*Department of Physics, Hanyang University, Seoul, 133-791, Korea*
[2]*Convergence Technology Division, Korea Advanced Nano Fab Center, Suwon 16229, Korea*
[3]*Joint Quantum Institute, University of Maryland, College Park, MD 20742, USA*
*These authors contributed equally to this work
[†]Correspondence should be addressed to yoonjw@hanyang.ac.kr

## Sample fabrication details

To fabricate the topological junction structure for validating our theoretical findings, we employ photolithography and E-beam lithography techniques. For the E-beam lithography process, Alignment markers and Chip lines necessary for component placement were produced using photo lithography. The Alignment marker consists of global and local keys, each with dimensions around 4 μm in width and 50 μm in length. Negative resist (DNR L300-D1, DONGJIN, KOR) with a thickness of 2 μm is coated on a Quartz wafer. Using an i-line stepper (NSR-2005i10C, Nikon), the patterns of the markers and keys engraved on the mask are exposed onto the wafer. The exposed chips undergo post-exposure bake (PEB) at 100°C before development. To enhance the adhesive properties of the metal on the pattern surface, surface treatment is conducted using O2 gas. Subsequently, Cr 10 nm and Au 200 nm are deposited using an E-beam evaporator (EI-5, ULVAC). Finally, the completion s of the align pattern for E-beam lithography is achieved through the metal lift-off process using acetone and IPA solutions.

The designed device consists of various periods and structural parameters, so we proceed with the meticulous E-beam lithography process as follows. In e-beam lithography, the area that can be exposed without moving the stage and by maximally deflecting the electron beam is called the working field. If the continuous pattern exceeds the working field, Stitching phenomena can occur. Considering these phenomena, the device area is set to 500 × 500 μm2. A SiN layer of 300 nm is deposited on the sample with completed align key pattern using PECVD (P500, Applied Materials). Next, after plasma treatment, a positive resist (ZEP520A) is coated to a thickness of 300 nm. Since a Quartz substrate is used in this process, an additional conductive layer step is required to prevent the charging effect caused by the electron beam. As a conductive layer, ESpacer 300Z is applied over the resist to improve the effects of charging. The photonics lattice pattern is exposed using E-beam lithography (JBX-9300, JEOL). During the exposure process with the electron beam, uneven patterns can form due to scattering among electrons, depending on the shape and density of the pattern. To prevent this phenomenon, Proximity Effect Correction (PEC) functionality is applied. PEC refers to the shot modulation capability that allows different electron beam energy to be applied to different areas of the pattern, thereby forming a structure with uniform linewidths. The process conditions applied in this E-beam lithography process are an acceleration voltage of 100 kV, electron beam current of 300 pA, and Dose of 150 μC/cm².

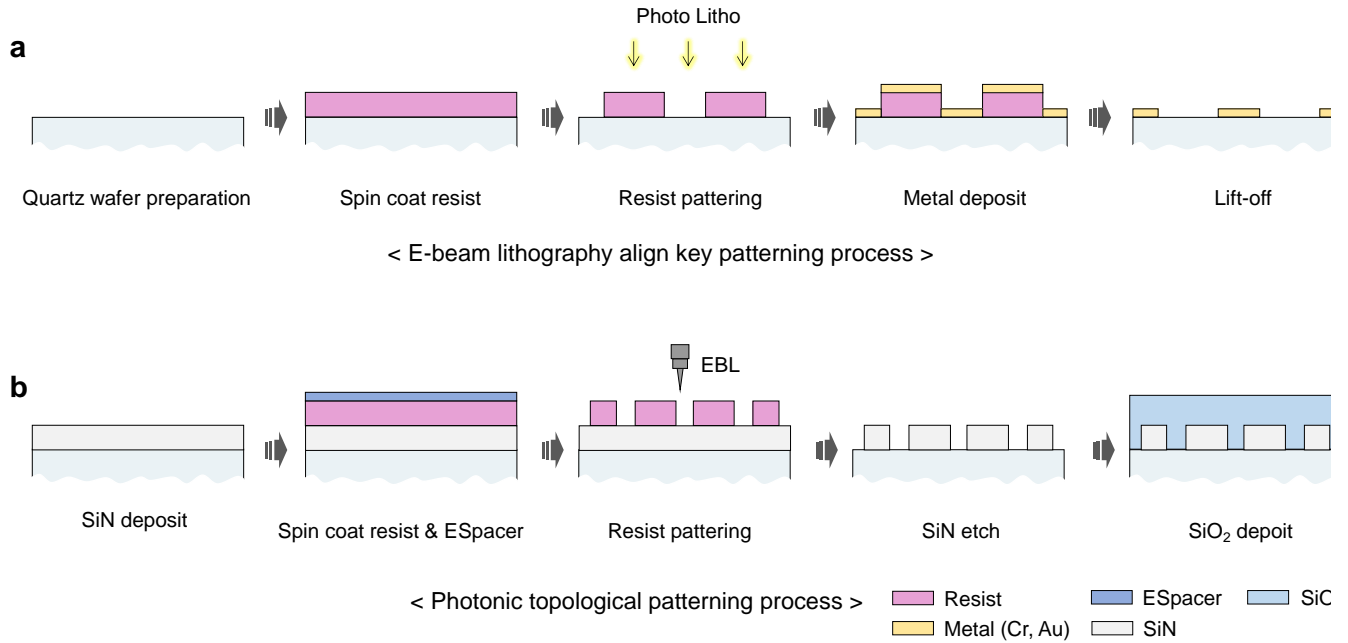

**Fig. S1 Step-by-step flow of the fabrication process. a** Align key pattern process for E-beam lithography. **b** Fabrication process of topological guided-mode resonance structures.

The resist patterns formed by e-beam lithography are etched using an Inductively Coupled Plasma Dry Etcher (Multiplex ICP, Oxford). Given the variety in periods and linewidths of the designed structures, there is a concern for micro/macro loading effects during the dry etching process. The micro/macro loading effect refers to the phenomenon where etch rates vary depending on the linewidth and density of the pattern, respectively. This effect is related to the removal of by-products, and to improve the micro/macro loading effect, the process should be conducted at low pressure. However, continuously lowering the pressure to improve the loading effect can lead to inadequate plasma formation. The ICP mode, which applies RF Bias to both the top and bottom of the equipment, has a higher plasma density at low pressures compared to RIE mode. This allows for the improvement of micro/macro loading effects that can occur in dry etching equipment, resulting in structures with consistent periods and linewidths, and uniform etch depths. The conditions for this etching process involved injecting process gases SF6 at 45 sccm and O2 at 5 sccm, with a pressure of 7 mTorr, applied power of 2 kW, and a bias of 50 V.